\documentclass[12pt]{article}
\usepackage{graphicx}
\graphicspath {./figs/}
\usepackage{times}
\usepackage{geometry}
\usepackage[utf8]{inputenc}
\usepackage{enumitem,amssymb}
\usepackage{natbib}
\usepackage{soul}
\usepackage[colorlinks=true,citecolor=blue,urlcolor=cyan]{hyperref}
\newcommand\degree{\degr}
\newcommand\degrees\degree

\newcommand\jwst{\em JWST}
\newcommand\hst{\em HST}

\DeclareSymbolFont{UPM}{U}{eur}{m}{n}
\DeclareMathSymbol{\umu}{0}{UPM}{"16}
\let\oldumu=\umu
\renewcommand\umu{\ifmmode\oldumu\else\math{\oldumu}\fi}
\newcommand\micro{\umu}
\newcommand\micron{\micro m}
\newcommand\microns \micron

\let\oldsim=\sim
\renewcommand\sim{\ifmmode\oldsim\else\math{\oldsim}\fi}
\let\oldpm=\pm
\renewcommand\pm{\ifmmode\oldpm\else\math{\oldpm}\fi}
\newcommand\by{\ifmmode\times\else\math{\times}\fi}

\newbox{\wdbox}
\renewcommand\c{\setbox\wdbox=\hbox{,}\hspace{\wd\wdbox}}
\renewcommand\i{\setbox\wdbox=\hbox{i}\hspace{\wd\wdbox}}

\newcount\timect
\newcount\hourct
\newcount\minct
\newcommand\now{\timect=\time \divide\timect by 60
         \hourct=\timect \multiply\hourct by 60
         \minct=\time \advance\minct by -\hourct
         \number\timect:\ifnum \minct < 10 0\fi\number\minct}

\catcode`@=11

\newcommand\comment[1]{}
\newcommand\commenton{\catcode`\%=14}
\newcommand\commentoff{\catcode`\%=12}

\renewcommand\math[1]{$#1$}
\newcommand\mathshifton{\catcode`\$=3}
\newcommand\mathshiftoff{\catcode`\$=12}

\comment{the backslash is necessary}

\comment{alignment tab}

\let\atab=&
\newcommand\atabon{\catcode`\&=4}
\newcommand\ataboff{\catcode`\&=12}

\let\oldmsp=\sp
\let\oldmsb=\sb
\def\sp#1{\ifmmode
           \oldmsp{#1}%
         \else\strut\raise.85ex\hbox{\scriptsize #1}\fi}
\def\sb#1{\ifmmode
           \oldmsb{#1}%
         \else\strut\raise-.54ex\hbox{\scriptsize #1}\fi}
\newbox\@sp
\newbox\@sb
\def\sbp#1#2{\ifmmode%
           \oldmsb{#1}\oldmsp{#2}%
         \else
           \setbox\@sb=\hbox{\sb{#1}}%
           \setbox\@sp=\hbox{\sp{#2}}%
           \rlap{\copy\@sb}\copy\@sp
           \ifdim \wd\@sb >\wd\@sp
             \hskip -\wd\@sp \hskip \wd\@sb
           \fi
        \fi}
\def\msp#1{\ifmmode
           \oldmsp{#1}
         \else \math{\oldmsp{#1}}\fi}
\def\msb#1{\ifmmode
           \oldmsb{#1}
         \else \math{\oldmsb{#1}}\fi}
\def\supon{\catcode`\^=7}
\def\supoff{\catcode`\^=12}
\def\subon{\catcode`\_=8}
\def\suboff{\catcode`\_=12}
\def\supsubon{\supon \subon}
\def\supsuboff{\supoff \suboff}

\newcommand\actcharon{\catcode`\~=13}
\newcommand\actcharoff{\catcode`\~=12}

\newcommand\paramon{\catcode`\#=6}
\newcommand\paramoff{\catcode`\#=12}

\comment{And now to turn us totally on and off...}

\newcommand\reservedcharson{\commenton \mathshifton \atabon \supsubon \actcharon
	\paramon}

\newcommand\reservedcharsoff{\commentoff \mathshiftoff \ataboff
	\supsuboff \actcharoff \paramoff}

\catcode`@=12
\reservedcharsoff

\reservedcharson

\comment{ Must have ONLY ONE of these... trust these macros, they work

}

\newcommand{\squishlist}{
 \begin{list}{$\bullet$}
  { \setlength{\itemsep}{1pt}
     \setlength{\parsep}{0pt}
     \setlength{\topsep}{3pt}
     \setlength{\partopsep}{0pt}
     \setlength{\leftmargin}{2.0em}
     \setlength{\labelwidth}{1.5em}
     \setlength{\labelsep}{0.5em} } }

\newcommand{\squishend}{
  \end{list}  }

\reservedcharson
\actcharon

\usepackage{tcolorbox}
\reservedcharson
\actcharon
\usepackage{bibentry}
\usepackage{astjnlabbrev}
\usepackage{wrapfig}
\usepackage{fancyhdr}
\usepackage{pdfpages}
 
\begin{document}

\includepdf[pages=-]{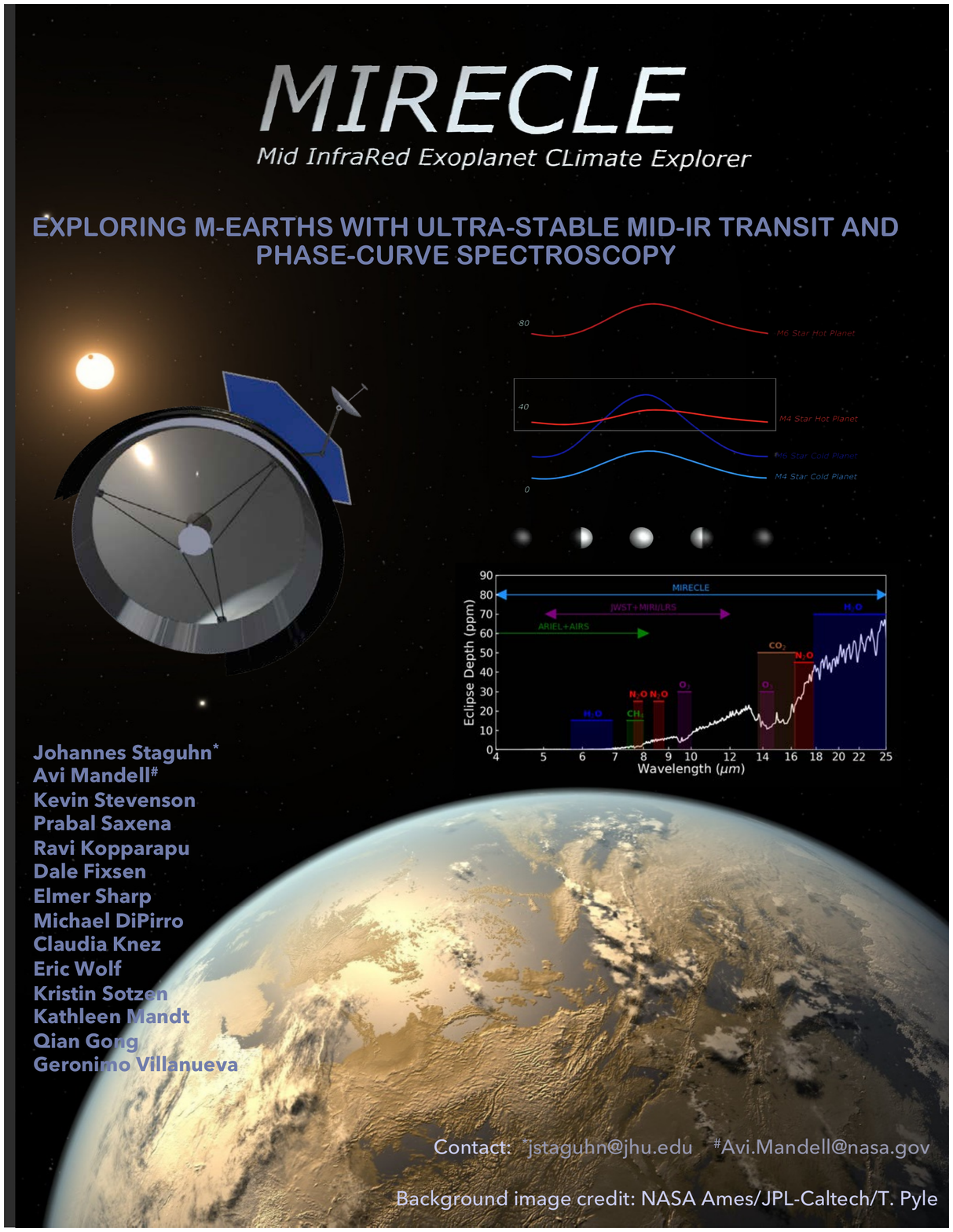}

\section{Key Science Goals \& Objectives}
\vspace*{-0.5\baselineskip}

Since before the discovery of the first confirmed exoplanets, the science community (and the broader public) has dreamt of discovering the first habitable and even inhabited planet around a nearby star.  The RV and transiting exoplanet discovery revolution, spearheaded by the Kepler and TESS detection missions and supporting efforts from ground-based surveys, are now discovering potentially rocky worlds around the nearest cool stars.  With the launch of the James Webb Space Telescope ({\jwst}) and subsequently the ESA/ARIEL mission, we will enter the era of comparative exoplanetology and take a significant step towards characterizing potentially-habitable worlds beyond our Solar System -- with the ultimate goal of detecting the first signs of life beyond Earth.  

Due to the nature of their instruments and wavelength coverage, {\jwst} and ARIEL will be limited in their study of habitable zone (HZ) planets.
\ul{We therefore recommend the prioritization of a stand-alone MIR mission for launch in the 2020's to characterize the structures and compositions of exoplanets orbiting the nearest cool stars.}

This White Paper presents a mission concept called \textbf{MIRECLE - the Mid-InfraRed Exoplanet CLimate Explorer}.  With a moderately sized aperture of 2 meters, broad wavelength coverage (4 -- 25 \microns), and next generation instruments, MIRECLE will be capable of efficiently characterizing a statistically significant sample of terrestrial planets, many of which will be in their host stars’ habitable zones. 
Spectroscopic characterization of terrestrial atmospheres will provide constraints for the distribution of planets with tenuous vs. substantial atmospheres, on the inner and outer edges of the habitable zone, and climate models to assess the potential for habitability. For the few brightest targets, the detection of specific combinations of molecules would provide evidence of biosignatures.
For all other targets, this comprehensive survey would filter out the airless, desiccated, or lifeless worlds, thus providing a subset of potentially-habitable worlds ready for in-depth atmospheric characterization using a larger aperture telescope.

\begin{tcolorbox}
\textbf{MIRECLE will:
\squishlist{}
\item Use a high precision calibration scheme for a Mid-IR Exoplanet Spectrometer, operating between 4 -- 25 \microns, with unprecedented photon-noise dominated performance (better than 5 ppm flux measurement stability) over multi-day timescales, utilizing an existing Transition Edge Sensor (TES) detector array and a self-calibration system.
    \item Enable high-precision mid-IR transmission, emission, and phase curve spectroscopy for the characterization of exoplanets, with the goal of enabling the detection of biosignatures in rocky planets around the nearest M dwarfs.
\squishend{}}
\end{tcolorbox}

\vspace*{-1.0\baselineskip}
\subsection{The Cool Star Advantage for Rocky Planet Characterization}
\vspace*{-0.5\baselineskip}

One of the early insights from transit discoveries was the concept of the ``M-star advantage'' for discovering and characterizing rocky planets.  M-dwarf stars, and cooler stars in general, provide an advantage in observing smaller planets due to the smaller stellar radius that results in a larger transit depth compared with larger stars.  Additionally, the demographics of planet occurrence mapped out by the Kepler mission reveal that the number of small planets per star increases for cooler stars (Dressing \& Charbonneau 2015). Finally, the overall stellar number density increases with decreasing stellar mass, which results in an increase in the number of potential nearby host stars. These factors make M dwarfs and even ultra-cool L dwarfs the prime candidates for hosting the first rocky exoplanets to be discovered and eventually characterized. Even more exciting is the fact that rocky planets in the HZ of cool stars have relatively short orbital periods (4 - 20 days), making them amenable to primary transit, secondary eclipse, and even phase curve measurements.  

There have been a wealth of studies trying to understand the composition, climate, and potential habitability of rocky and temperate M-dwarf planets, colloquially called ``M-Earths'' (see Shields et al. 2016 for a recent review), and there is considerable uncertainty about the likelihood that these planets could actually support a viable biosphere over significant time periods.  However, there is no doubt that the opportunity to study cool rocky planets beyond our solar system for the first time would be invaluable to our understanding of planetary environments in general.



\vspace*{-1.0\baselineskip}
\subsection{The Mid-IR Advantage for Climate Studies of Temperate M-Earths}
\vspace*{-0.5\baselineskip}

In the mid-infrared, the main observable is a planet's thermal emission spectrum, which is measured using the secondary eclipse technique for a planet's dayside or the phase curve technique for all other longitudes.  Both methods probe a planet's atmospheric composition and thermal structure, which are critical towards assessing its habitability.  The latter method also constrains the transport of heat from the dayside to the nightside and informs on the presence and distribution of aerosols (defined as clouds or hazes).

\begin{figure}[t]
\centering
\includegraphics[width=0.75\linewidth]{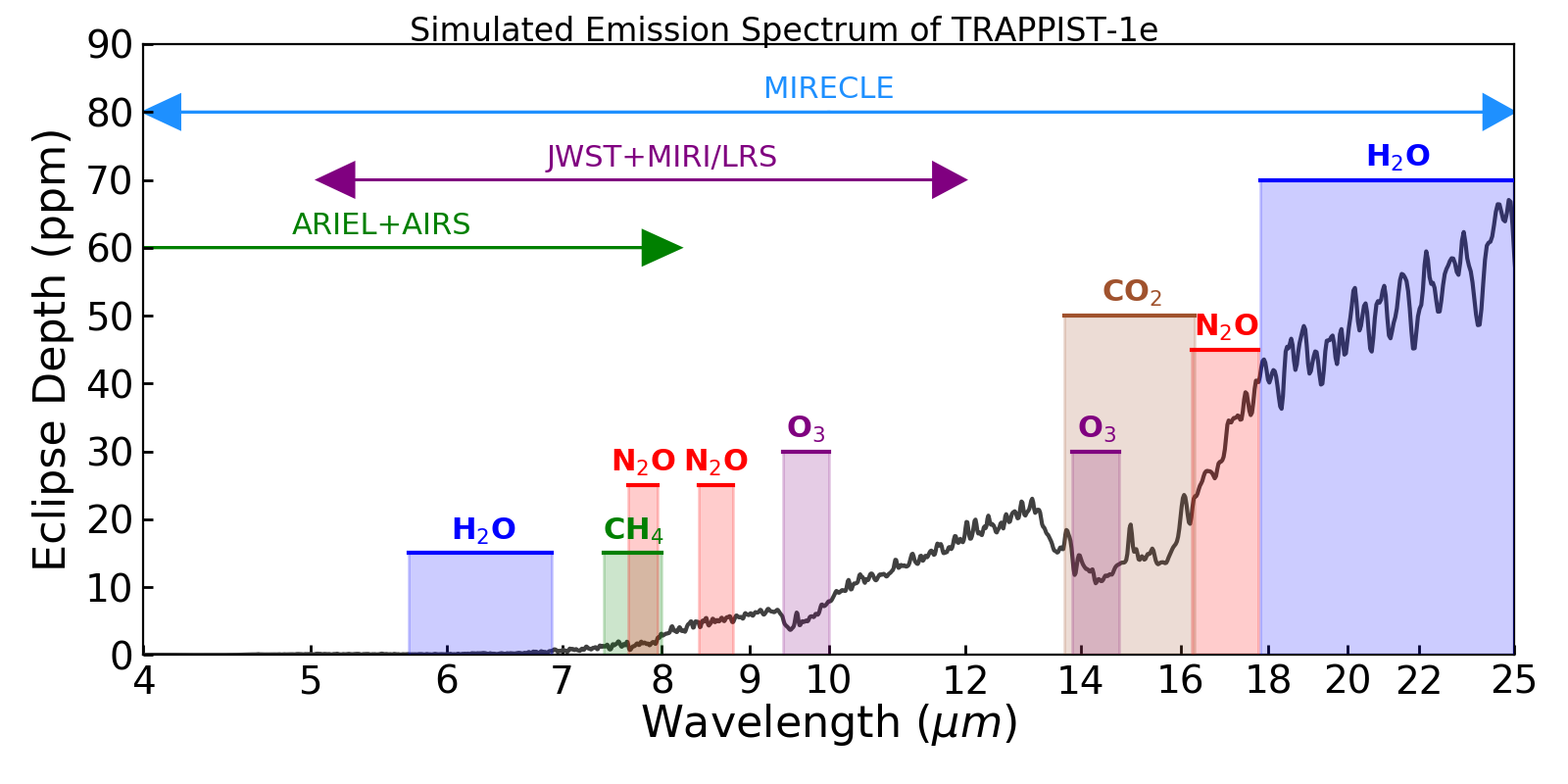}
\vspace*{-1.0\baselineskip}
\caption{\label{fig:spectrum}{\small 
Simulated emission spectrum of a HZ planet (TRAPPIST-1e) with the corresponding mid-IR wavelength coverage of future missions.  Only MIRECLE has the mid-IR wavelength coverage to adequately constrain the composition and thermal structure of HZ planets through emission spectroscopy.
}}
\end{figure}

For terrestrial exoplanets orbiting within the habitable zones of M-dwarf stars, the planet-to-star flux contrast ratio becomes favorable at wavelengths $ > 8 ~\mu$m (see Figure \ref{fig:spectrum}).
Accounting for the decrease in photon count rate at longer wavelengths, the signal-to-noise ratio is favorable out to $\sim25~\mu$m. Within this wavelength range, there are prominent absorption features due to $CH_{4}$, $CO_{2}$, $O_{3}$, $NH_{3}$, $N_{2}O$, and $SO_{2}$, as well as the $H_{2}O$ vapor continuum. 
These features can readily distinguish a wet, Earth-like planet from a dry, Venus-like planet with a dense $CO_{2}$ atmosphere and a Mars-like planet with a thin $CO_{2}$ atmosphere. The strong ozone band at 9.7~$\mu$m allows for detecting the presence of molecular oxygen in the atmosphere, which is a powerful biosignature when combined with other out-of-equilibrium molecular species (such as $CH_{4}$ and/or $N_2O$). 

In addition to measuring planetary emission, mid-infrared observations can take full advantage of a planet's transmission spectrum across a wide wavelength range. Spectra from 5 - 25 $\mu$m are less sensitive to the high-altitude aerosols that tend to obscure transmission spectra at shorter wavelengths, and provide access to different molecular species than NIR wavelengths.  Furthermore, transmission spectroscopy probes pressure levels above Earth's tropopause, thus mitigating the effects of water clouds.


Specific questions that a mid-IR mission could address include:
\squishlist{}
\item What fraction of temperate, terrestrial planets orbiting M dwarfs have appreciable atmospheres?
\item What physical properties (e.g. planet mass, stellar flux, etc.) set the habitability of terrestrial M-dwarf planets at the inner and outer edges of the habitable zone?
\item What is the atmospheric oxidation state for terrestrial M-dwarf planets, and how does it vary with measured physical properties?
\squishend{}
JWST will likely be able to answer some of these question for a small handful of planets orbiting mid-to-late M dwarfs; however, it is unknown if JWST’s instruments will have the precision to answer these questions for the early-M dwarfs and it is unlikely that JWST will have the available telescope time to study a statistically significant sample size.  In many respects, MIRECLE can serve as a reconnaissance mission for future flagships such as the Origins Space Telescope.

\vspace*{-1.0\baselineskip}
\subsection{Key Science Goal: Revealing the Nature of Habitable Rocky Planets Around Cool Stars}
\subsubsection{Measuring Global Weather Patterns via Spectroscopic Phase Curves}
\vspace*{-0.5\baselineskip}

Similar to the hot Jupiter phase-curve observations conducted with {\hst} and eventually {\jwst}, a mid-IR telescope would be able to measure phase-resolved thermal emission (i.e., spectroscopic phase curves) of temperate planets orbiting M-dwarf stars. Because these close-in planets are expected to be tidally locked (i.e. synchronously rotating), rotation periods are expected to be on the order of days to weeks -- much longer than the rotation periods of many solar system planets. These properties strongly influence a planet’s atmospheric circulation (e.g., Carone et al., 2018).  For slower rotators ($P>20$ days), the circulation of terrestrial M-dwarf planets can exhibit weak radial flow from dayside to nightside and the so-called ``eyeball'' climate state, in which a liquid water ocean only exists near the sub-stellar point (e.g., Pierrehumbert, 2010; Angerhausen et al., 2013).  Comparatively faster rotators ($P\sim1$ day) can exhibit broad upper atmosphere super-rotation, or high-latitude, super-rotating jets with banding (Carone et al., 2018). If the planet’s orbit is eccentric, pseudo-locking of its orbit and rotation may result in weak seasonal climatic cycles (e.g., Driscoll \& Barnes, 2015).  Each of these circulation patterns (i.e. weather) influence the atmosphere's distribution of heat and the formation and transport of aerosols, all of which shape thermal phase curves.

Mid-IR phase curve observations of terrestrial M-dwarf planets can be used to constrain a wealth of properties. First, they can be used to distinguish between planets with and without substantial atmospheres, and subsequently identify the presence of hemispheric-wide clouds (Yang et al., 2013).  Second, phase curve observations can constrain the planet’s rotation rate --- the morphology of thermal phase curves can differentiate synchronous rotators from those in spin-orbit resonances (Wang et al., 2014). This is also applicable to eccentric planets (e.g., Boutle et al., 2017). Third, the amplitude and morphology of phase curve observations can help distinguish between different climatic states (Wolf, 2017; Kopparapu et al., 2017).  Climate modeling results show that, for terrestrial planets around M dwarfs, there is a clear distinction between runaway and non-runaway greenhouse climate states. 

Phase curve observations can also be conducted for non-transiting planets, where they can detect the presence of an atmosphere, distinguish between different atmospheric compositions (i.e., dry planets vs. ``aquaplanets''), and differentiate between different rotation rates (Turbet et al., 2016).  Importantly, thermal phase curves of non-transiting planets can be used to measure the planet’s inclination, as has been demonstrated for hot Jupiters (Crossfield et al., 2010).

\begin{figure}[t]
\centering
\includegraphics[width=0.48\linewidth]{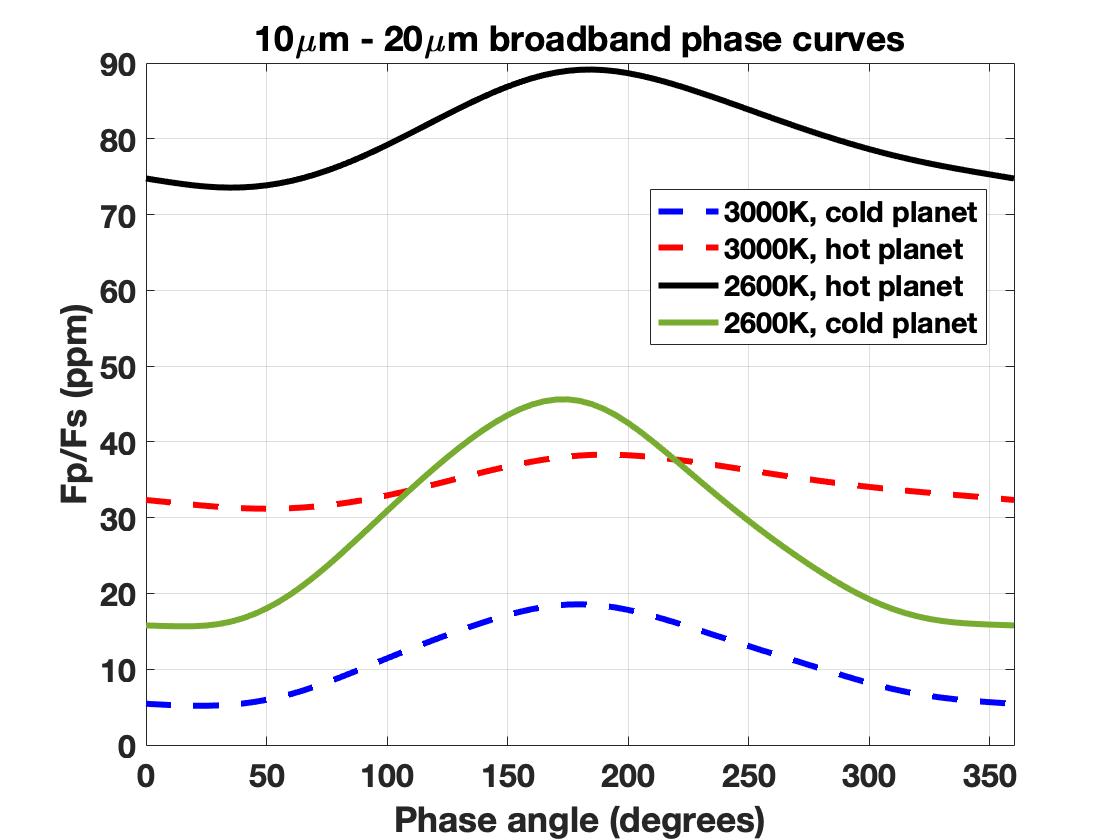}
\includegraphics[width=0.51\linewidth]{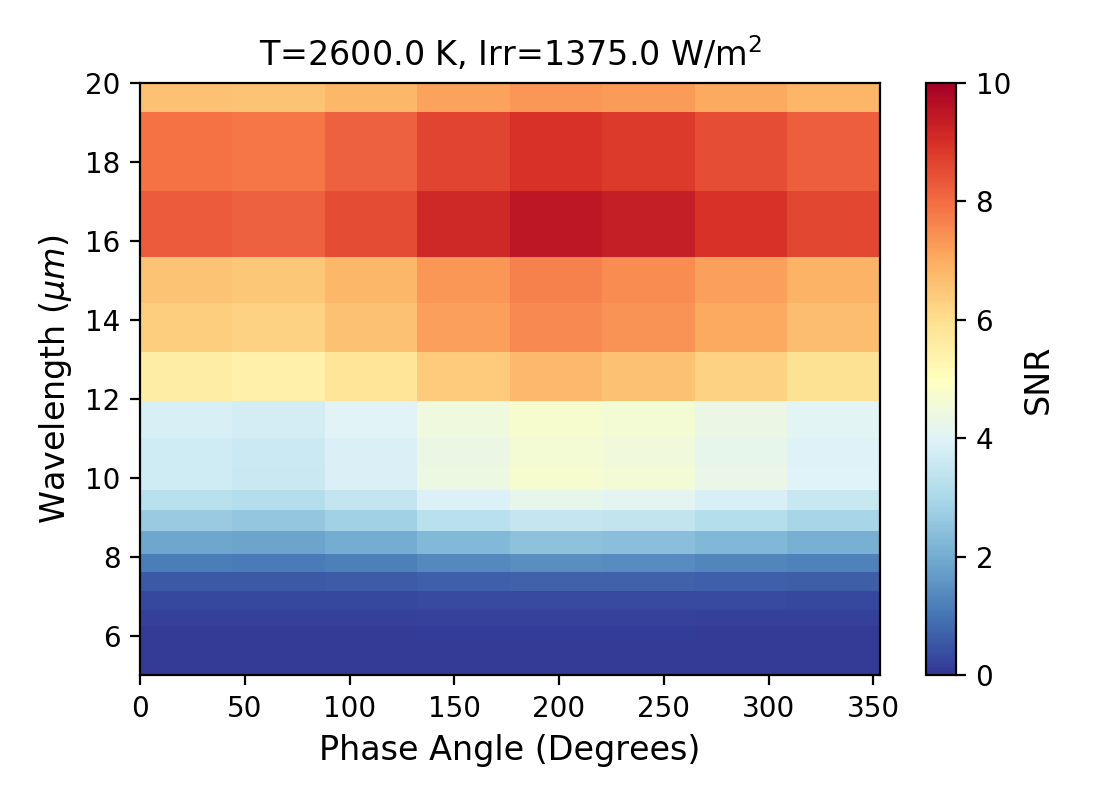}
\vspace*{-1\baselineskip}
\caption{\label{fig:phasecurve}{\small 
{\bf Left:} Broadband phase curves of model terrestrial planets orbiting M4 (3000~K) and M6 (2600~K) stars. A phase angle of $0^{\circ}$ corresponds to the anti-stellar point (transit), and $180^{\circ}$ corresponds to the sub-stellar point (eclipse). Differentiating the inner and outer edges of the HZ (hot vs cold planet) is feasible for mid-to-late M dwarfs with an ultra-stable instrument and an observatory that can observe continuously for several weeks.
{\bf Right:} Wavelength- and time-dependent signal-to-noise ratio (SNR) estimates for a hot ($\sim$340~K) planet orbiting a 2600~K star.  The simulation assumes 20 phase curves with MIRECLE and a K\sb{mag}~=~8 star. It is clear that targeting the 10--20 {\micron} region is required to detect emission from the planet.
}}
\end{figure}

Phase curve observations require continuous observations of up to several weeks and a wavelength range of 5 -- 20+~{\microns}.  For comparison, JWST’s spectroscopic capabilities are limited to $<12$~{\microns} (for time-series observations) and $<48$ hours in duration (before having to re-point). Additionally, such measurements will require support and input from the broader science community to explore critical questions including the nature of stellar variability of M-dwarfs, and the role of uncertainty in planetary radii influence the signal of potential spectral signatures.

\vspace*{-0.5\baselineskip}
\subsubsection{Extending Transmission Spectroscopy of M-Earths to MIR Wavelengths}
\vspace*{-0.5\baselineskip}
Planets orbiting in the HZs of M dwarf stars have relatively short orbital periods (5 - 40 days) and are theorized to receive higher levels of UV flux.  Extreme UV flux can enhance atmospheric escape to the point of yielding only a tenuous atmosphere.  M dwarfs tend to be most active shortly after formation ($<1$ Gyr), thus potentially impacting planets that formed in-situ.  Planets that have migrated into the HZ after 1 Gyr are more likely to retain appreciable atmospheres.  Also, the secondary atmospheres of impacted (e.g. in-situ formation) planets could be replenished from infalling cometary material.  It is therefore unknown whether temperate, terrestrial planets orbiting M dwarfs have appreciable atmospheres.  Understanding what fraction of temperate terrestrial planets with short orbital periods have substantial atmospheres can inform us about the formation and migration of these systems and place our own planet into a broader context.

The easiest way to determine if a planet has an atmosphere is to search for spectral modulation in the transmission spectrum.  The most dominant spectral features are anticipated to be CO\sb{2} at 2.7, 4.3, and 15 {\microns}. The instrument would require $<5$ ppm precision to measure CO\sb{2} in an Earth-size planet transiting an M2 star.  Thus, the telescope pointing and instrument temperature would need to be extremely stable.  Furthermore, the telescope would require a large instantaneous field of regard to obtain a large number of primary transits within its mission lifetime.

\begin{figure}[t]
\centering
\includegraphics[width=0.49\linewidth]{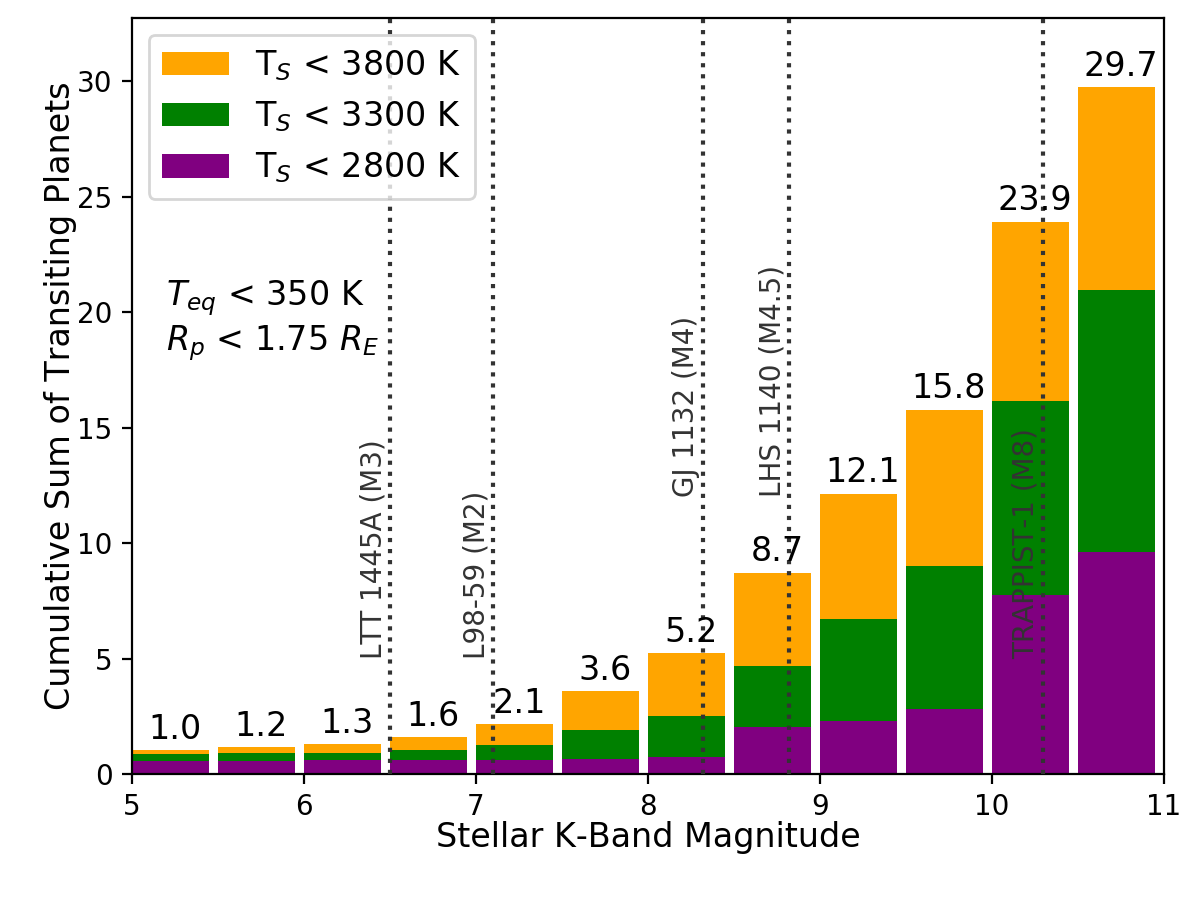}
\includegraphics[width=0.5\linewidth]{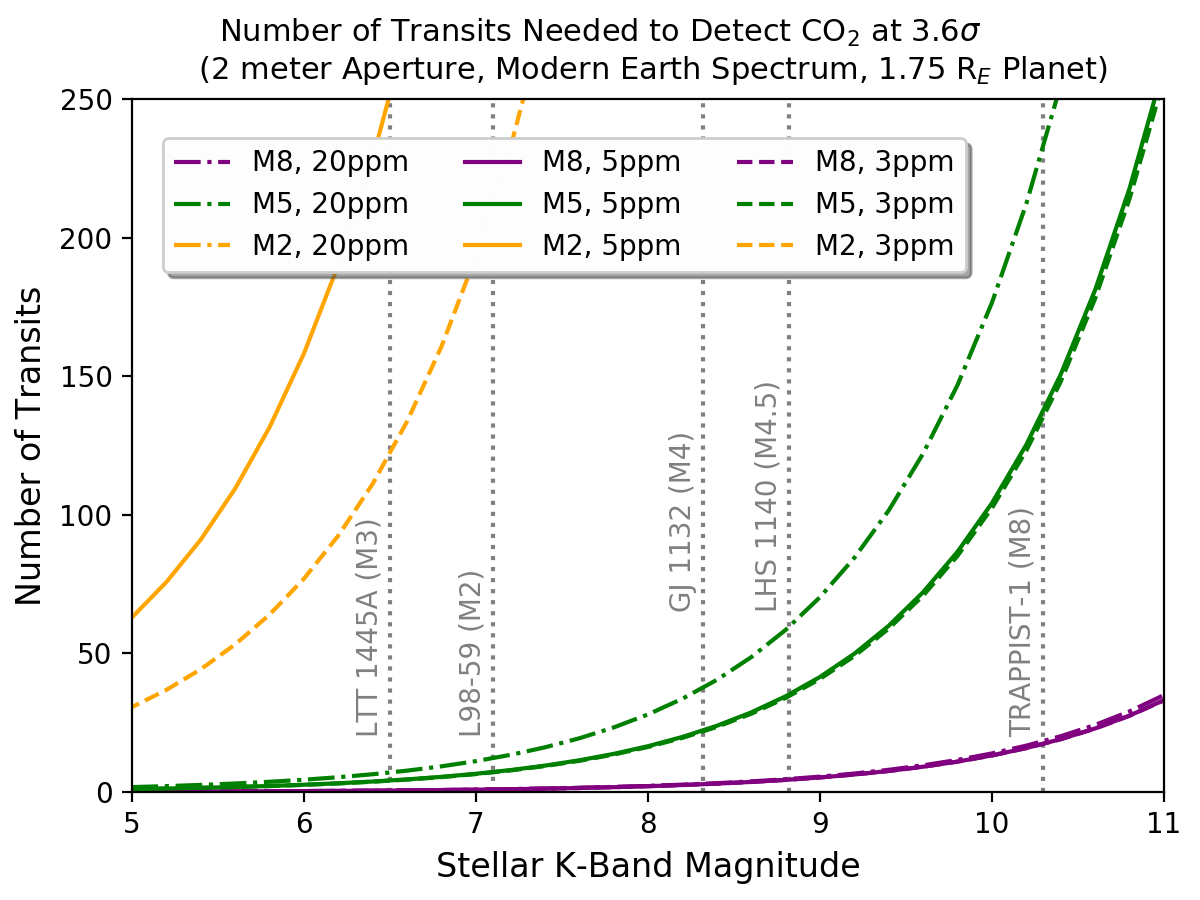}
\vspace*{-1\baselineskip}
\caption{\label{fig:co2}{\small 
{\bf Left:} Anticipated exoplanet yields from the TESS and SPECULOOS surveys (Barclay et al., 2018; Delrez et al., 2018).  Yields are limited to terrestrial, habitable-zone planets transiting M dwarfs.
Additional planets are likely to be discovered through other surveys or during extended missions.
{\bf Right:} The number of transits needed to confirm the presence of an atmosphere (through the detection of CO\sb{2}) depends strongly on the stellar magnitude, stellar type (M2 -- M8 shown), and instrument noise floor (3 -- 20 ppm shown).  A JWST-like, 20~ppm noise floor is insufficient to detect spectral modulation for the planets orbiting LTT~1445A and L98-59, both early-to-mid M dwarfs.  For these and similar systems soon to be discovered by TESS, an instrument noise floor of $<5$ ppm is necessary to distinguish between tenuous, clear, or cloudy atmospheres.
}}
\end{figure}

JWST will likely be able to determine this for a small number of planets orbiting mid-to-late M dwarfs; however, it is unknown if JWST’s instruments will have the precision to answer this question for the early-M dwarfs where the CO\sb{2} feature size is predicted to be $3 - 12$ ppm (see Figure~\ref{fig:co2}) and it is unlikely that JWST will have the available telescope time to study a statistically significant sample size.  

A dedicated MIR mission could explore a broad region of parameter space encompassing planet size, temperature, semi-major axis, and stellar type to distinguish between tenuous, clear, and cloudy atmospheres. These data would enable the identification of trends that could be used to inform planet formation models and predict which planets are best suited for an in-depth search for signs of life.

\subsection{Auxiliary Exoplanet Science: Solving Key Questions in Planet Formation and Planetary Structure}

{\bf Bridging the Gap Between Hot Jupiters and Solar System Planets:}
The last decade of exoplanet science has shown there is a broad continuum of worlds from sub-Earths to Super-Jupiters around Sun-like stars. For giant planets with hydrogen-dominated atmospheres ($>10$ Earth masses), missions such as JWST and ARIEL will target warm planets ($>600$ K) in thermal emission. This will leave a gap in our understanding between these objects and the solar system giant planets ($<200$ K).  Spectroscopic mid-IR observations will enable transformational science in the chemistry and structure of temperate giant planets around stars from types A to M. In these cooler atmospheres, the abundances of CH\sb{4} and NH\sb{3} probe metallicity, non-equilibrium chemistry, and the strength of vertical mixing (Zahnle \& Marley, 2014).  A mid-IR telescope could probe the rich photochemistry expected for these atmospheres, via production of HCN, C\sb{2}H\sb{2}, and C\sb{2}H\sb{4} from methane, as is seen in Jupiter beyond 5 {\microns}. The population of giant planets that could be probed in thermal emission (down to $\sim250$ K) will be cool enough to, for the first time, serve as a bridge between the extrasolar planet population and our own solar system.

{\bf Understanding the super-Earth/sub-Neptune boundary:}
The size of a planet is a directly-observable property that reveals clues about its history of formation and evolution.  Planets below 1.6 Earth radii are thought to be predominantly rocky (Rogers, 2015), but it’s not known if they formed that size or were once larger and subsequently lost their primordial atmospheres.  Recently, the Kepler observatory has shown us that there is a gap in the occurrence rate of planets between 1.5 and 2.0 Earth radii (Fulton et al., 2017).  This paucity of planets supports the idea that a sub-population of close-in sub-Neptune planets undergo a relatively quick process of atmospheric mass loss that leaves them with a rocky core measuring up to 1.6 Earth radii and a thin, high-mean-molecular-weight atmosphere (i.e. super Earths).  A large characterization survey of super-Earth and sub-Neptune size planets would address questions such as: (1) Why are super-Earths and sub-Neptunes the most common type of planet? (2) Why do the super-Earths typically reside in more highly-irradiated environments as compared to the sub-Neptunes? (3) What additional factors help distinguish super-Earths from sub-Neptunes? (4) Why are there so few planets in the “photoevaporation dessert”? Direct observations of their atmospheres will confirm theoretical predictions of the formation and evolution of super-Earths and sub-Neptunes.

{\bf Revealing the Nature of Clouds in Hot Jupiters:}
A mid-IR telescope can advance our understanding of the physics and chemistry of clouds, which are currently a major source of uncertainty in models of exoplanet atmospheres.  Since most clouds appear only as Rayleigh or gray scatterers in the optical and near-IR, we have no handle on the composition of clouds that we infer today in hot Jupiter atmospheres.  A number of cloud species expected to impact the spectra of these planets have known features in the mid-IR (e.g., Wakeford \& Sing, 2015).  Examples include: Al\sb{2}O\sb{3} (12 {\microns}), CaTiO\sb{3} (13 {\microns}, 22 {\microns}), and Mg\sb{X}SiO\sb{Y} (8 - 22 {\microns}).  {\jwst} cannot access these species using currently-supported time-series modes.  A mid-IR telescope would enable the unique identification of the clouds that impact the transmission spectra of hot Jupiter exoplanets.



\vspace*{-0.5\baselineskip}
\section{Technical Overview}
\vspace*{-0.5\baselineskip}



\subsection{Overview}
\vspace*{-0.5\baselineskip}
The MIRECLE observatory design is conventional with a 2m Cassegrain secondary and a typical dispersive element (either prism, or grating) to generate the spectrum. Since the telescope serves only as a light bucket (no need to image the unresolved star), minor large scale imperfections in the primary are not critical.

The proposed effort takes advantage of a new mid-IR detector and calibration technology, necessary to enable the detection of many important atmospheric lines, possibly including bio-signatures in planetary atmospheres. The required detector stability over many hours, needed for the detection of bio-signatures in the atmospheres of exoplanet through transit spectroscopy in the mid-IR, will only be possible through a very stable detector array plus an integrated ultra-stable calibration scheme as proposed here. We emphasize that the stability requirement for the detector is independent of the distance to the observed planet.

\begin{tcolorbox}[width = 15.0cm] 
\textbf{Instrument Key Attributes:
\squishlist{}
\item Photon-noise limited performance over the entire band (4 -- 25\microns)
\item High stability for phase curves and transit spectroscopy
\item Integrated high-precision calibration
\squishend{}}
\end{tcolorbox}

\vspace*{-0.5\baselineskip}
\subsection{Cryogenic Architecture of MIRECLE}
\vspace*{-0.5\baselineskip}

\begin{figure*}
\includegraphics[width=0.5\textwidth]{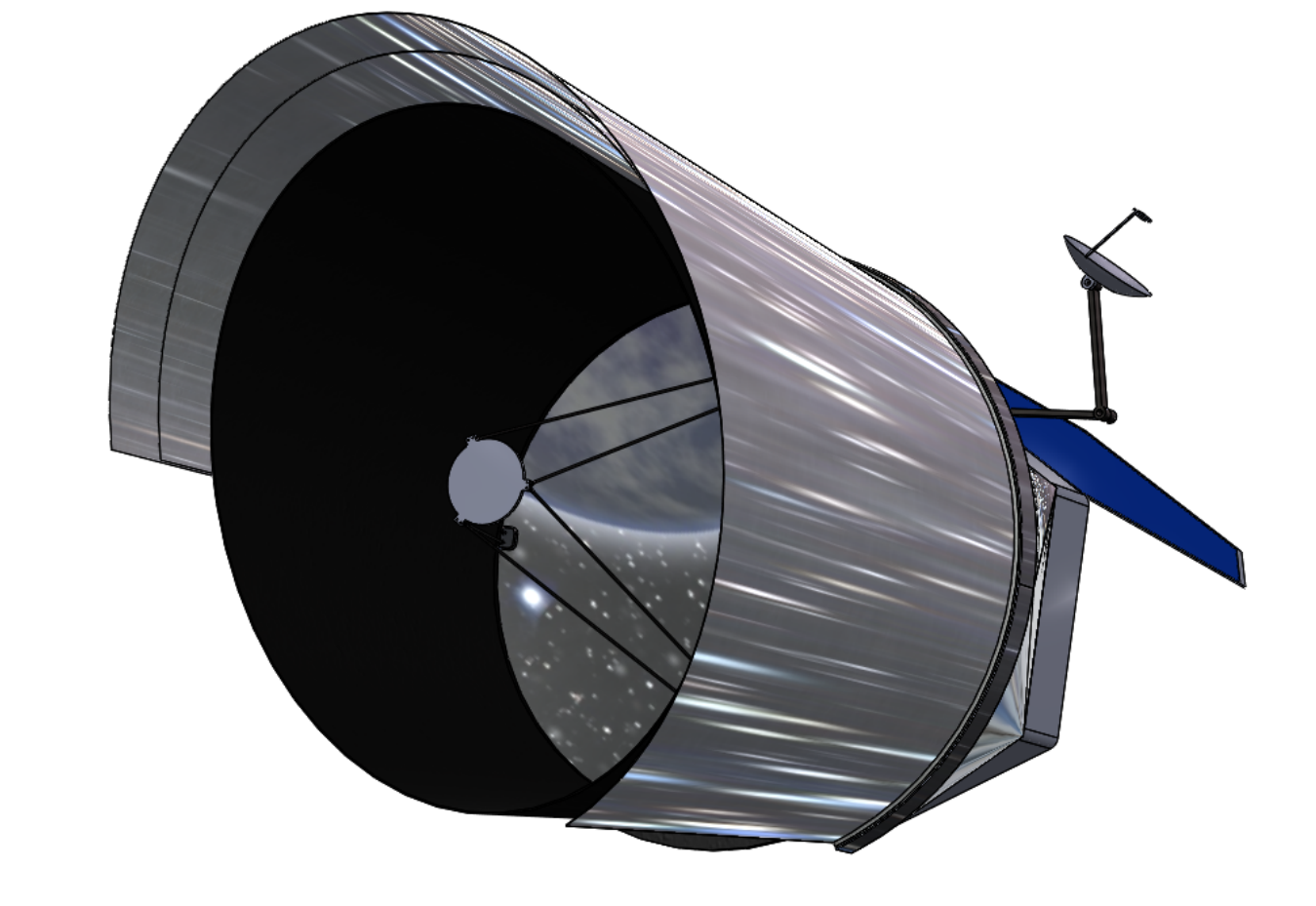}
\includegraphics[width=0.5\textwidth]{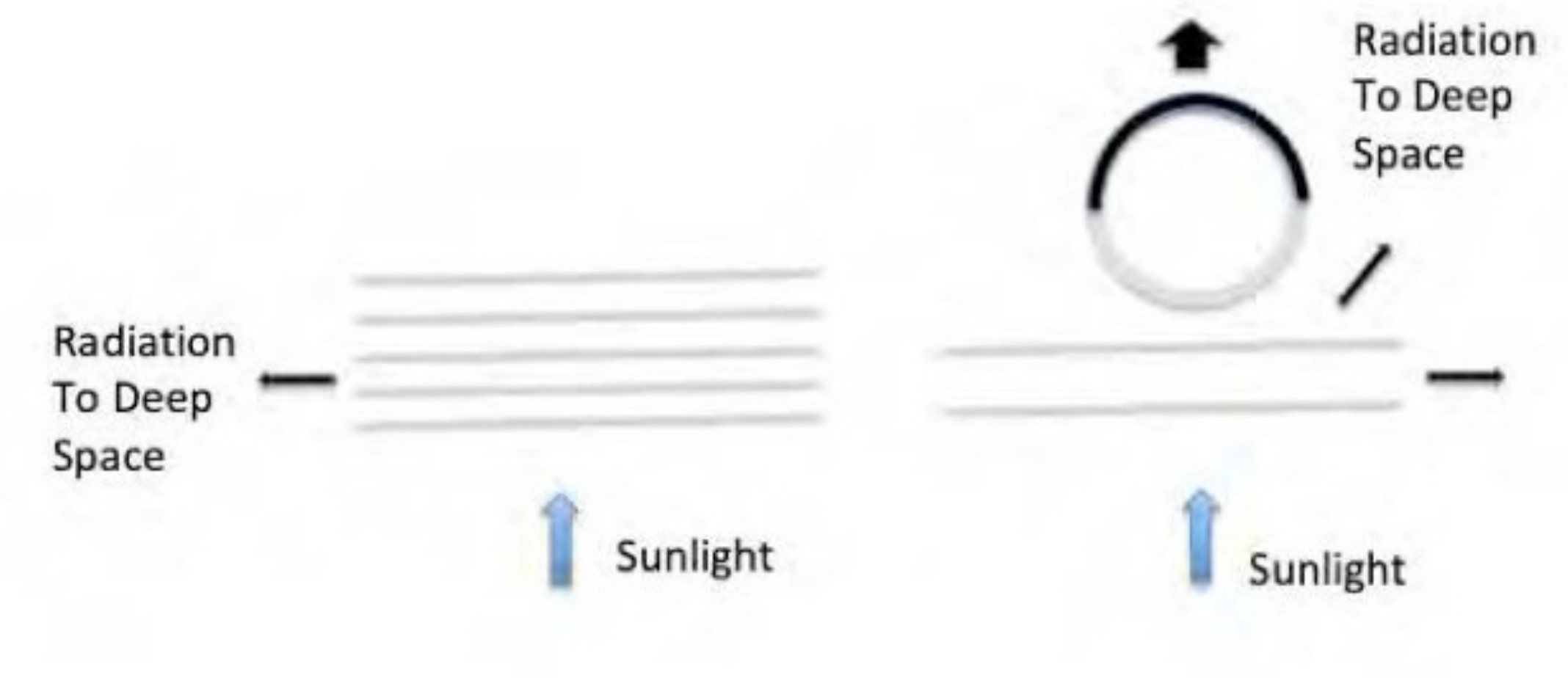}
\caption{A two-layer sunshield plus radiator can achieve the same radiative temperature as a five layer sunshield. A cutaway picture of MIRECLE is shown on the left.\label{fig:cryo}}
\end{figure*}

The payload will be most efficiently cooled by a combination of radiative and mechanical cooling. MIRACLE uses the Spitzer architecture (Figure~\ref{fig:cryo}), but with a 4.5 K cryocooler substituted for a helium dewar. Two stationary sunshields, a cylindrical barrel, and baffle protect the 2-m diameter telescope and instruments from sunlight, Earth-shine and Moon-shine.  The barrel is radiatively cooled by coating the deep-space facing side with high-emissivity black paint. In the case of Spitzer this surface reached 34 K.  Inside the barrel the telescope will be cooled by an upper stage of the 4.5 K cryocooler. The telescope and baffle will operate at 20-25 K, depending on the cryocooler technology chosen. (See below).  The 4.5 K stage will cool the instrument, and act as a heat sink for the 50 mK Adiabatic Demagnetization Refrigerator (ADR) which cools the Transition Edge Sensor (TES) detectors. Current technology to achieve the required cooling is at TRL 5 with technology advancement currently funded and/or planned to reach TRL 6 by 2023.

\vspace*{-0.5\baselineskip}
\subsection{Spacecraft Systems}
\vspace*{-0.5\baselineskip}

MIRECLE does not present any particular challenges for the spacecraft.  The power requirements are about 1 kW using a standard deployed solar array.  The thermal system can reject the heat using body mounted panels-no deployables. There will be minimal on-board data processing.  The data system will generate a data rate, which can easily be downlinked by Ka band transponders and a gimballed high-gain antenna.  Attitude control can be achieved to the level of 10 milli-arc-sec by standard gyros, star trackers and a telescope fine steering mirror.  The planned orbit at Sun-Earth L2 (SEL2) uses a reasonable amount of fuel ($\sim 100$ m/s delta V required over the mission lifetime) which represents about 10\% of the observatory mass.

\vspace*{-0.5\baselineskip}
\section{Technology Drivers}
\vspace*{-0.5\baselineskip}
\subsection{Introduction}
\vspace*{-0.5\baselineskip}
Many current mid-IR detectors are based on impurity band conduction (IBC) devices such as Si:As detectors. Charge trapping in these device leads to a time and exposure dependent response. As a result, this detector class is not expected to provide the required 5 ppm stability over several hours of integration (for transit measurements) or days (for phase curve observations) which is needed for reliable detection at the 5 ppm level. While efforts are under way to improve IBC detectors, it is unclear how far the performance can be improved. 
Here we describe an ultra-stable Mid-IR Array Spectrometer for Exoplanet Transit observations, including two key technological advances:
\begin{itemize}
    \vspace*{-0.3\baselineskip}
    \item A real-time calibration system that continually provides a stable illumination to re-calibrate the detector sensitivity 
    \vspace*{-0.5\baselineskip}
    \item The use of arrays of ultra-stable Transition Edge Sensor detectors (TES)
\end{itemize}
\vspace*{-0.3\baselineskip}

TESs have a very linear response and are intrinsically very stable and  the required sensitivity and dynamic range can be easily met with existing devices for space based mid-IR transit spectroscopy. No new detector developments are required, only the absorbers and resonant backshorts need to be optimized for the wavelength range of the instrument. The calibration system will utilize a highly-calibrated blackbody source to periodically correct for small drifts in the detector gain over long timescales.  Together, these technologies will help to overcome the current noise floor for time series measurements. 

\vspace*{-0.5\baselineskip}
\subsection{Calibration System}
\vspace*{-0.5\baselineskip}

 \textbf{The need for calibration on minute time scales:} To guarantee photon noise dominated performance, the spectrometer must remain stable for the duration of an observation, i.e. several hours for transits and days for phase observations.  TES detectors have already demonstrated stability for minute timescales, but the overall detector gain drifts over longer times.  We have designed a calibration system that leverages that to days. A simple tungsten filament is very stable if it is kept in a vacuum below 2500K so the changes in the filament itself (evaporation, oxidation etc.) are very slow. To stabilize the temperature we use a photo diode $0.5\mu$m detector with continuous feedback to maintain a stable output of $0.5\mu$m light from the black body calibrator. Modulating this signal reveals the gain of the detector. Further details are available in Staguhn et al., 2019.
  
\begin{figure*}
\includegraphics[width=0.61\textwidth]{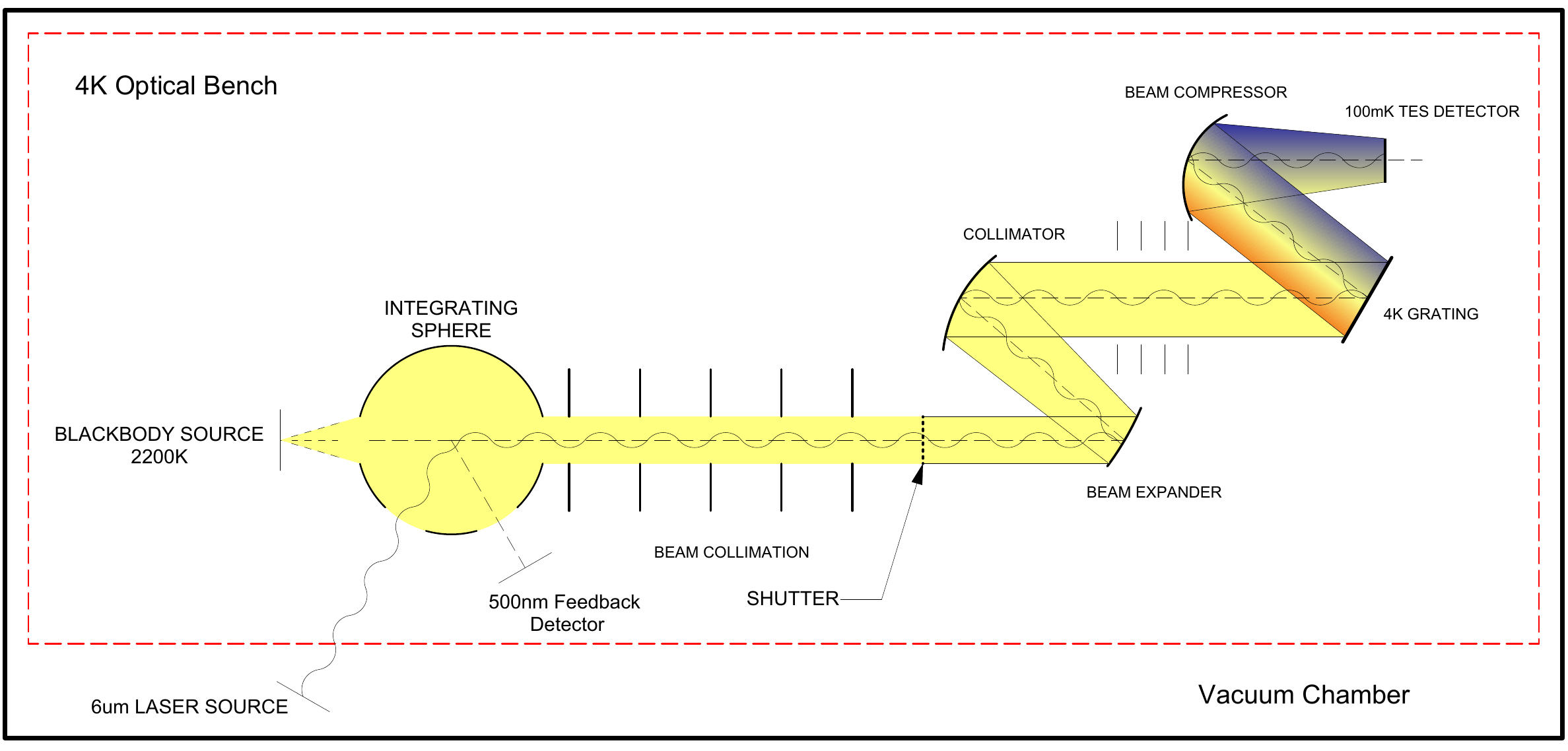}
\includegraphics[width=0.38\textwidth]{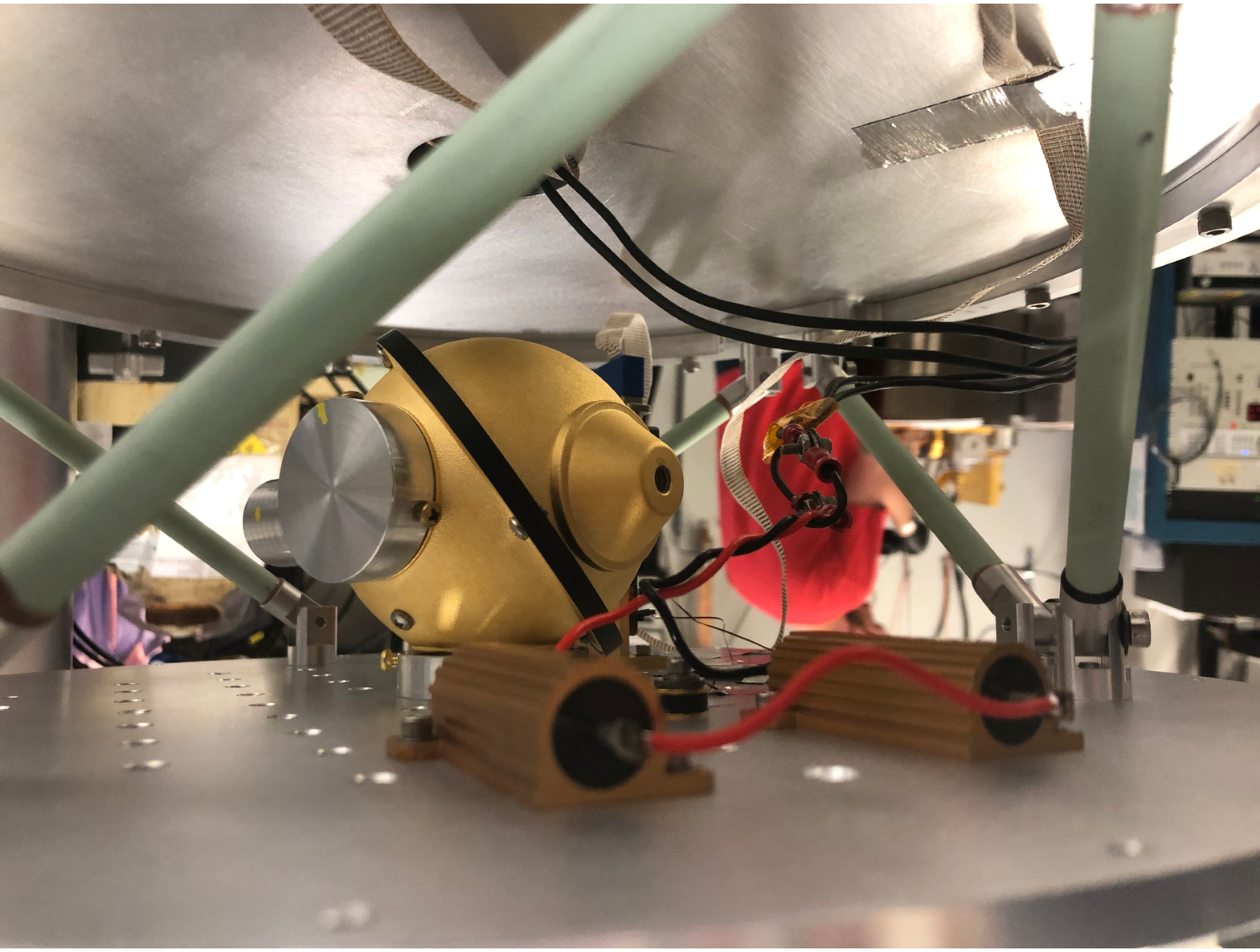}
\caption{\small
{\bf Left:} Schematic showing the laboratory experiment setup. The calibration system (blackbody source, integrating sphere, and the photo-diode for temperature monitoring) are all at 4 K together with the cold optics, mid-IR grating, and the TES detector array at 100 mK. The infrared laser, which substitutes for an observed spectral line feature, will be temperature controlled at the 50 K stage. Note that the optical coupling of the laser will not be very sensitive to temperature fluctuations and that the laser only serves as frequency calibration. Amplitude variations in the laser output do not affect the frequency  calibration. 
{\bf Right:} The integrating sphere was installed on an optics plate together with heaters and thermometers for tests of the PID active temperature control system. These tests are planned with existing funding for this year in the pre-proposal phase.\label{fig:labspectro}}
\end{figure*}

\textbf{Laboratory Demonstration:} To provide an adequate demonstration, both the spectral resolution and the stability over several hours needs to be demonstrated. A prototype instrument, named MIRASAT, has been partially funded through NASA's APRA program and is described in detail in Staguhn et al., 2019. The demonstration will be built in a modified lab dewar to maintain the detectors and the rest of the instrument at a cold and stable temperature. A sketch of the laboratory design is shown in Figure~\ref{fig:labspectro} ({\em left}).  Parts of the experiment have already been assembled (Figure \ref{fig:labspectro}, {\em right}) and first temperature control measurements of the calibration system have begun. 

\vspace*{-0.5\baselineskip}
\subsection{Cooling Technology}
\vspace*{-0.5\baselineskip}
Sensitive measurements at greater than 20{\microns} require a telescope temperature of $\sim10$K. Space instrument cooling to less than 30 K has advanced greatly over the last two decades to the point where using a 4.5 K cryocooler, rather than stored liquid helium is now the baseline.  The TRL 7 JWST/MIRI cryocooler is designed to cool a primary load at 6.2 K and an intercept load at 18 K. It is a three-stage pulse tube cryocooler to 18~K with a helium Joule Thompson loop to 6.2 K. The only change required to lower the temperature to 4.5 K is to reduce the return pressure from of the Joule Thompson loop. This can be achieved by upgrading the JT compressor used in the MIRI cooler.


Additionally, a key requirement for TES detector performance is an operating temperature of 100 mK. Cooling of detector arrays to 50 mK has already been demonstrated on Hitomi, with further improvements in mass and cooling power currently under development, extending this more efficient technology from TRL-4 to -6 (Tuttle, 2017). A three-stage ADR used on Hitomi produced 0.4~$\mu$W of cooling at 50 mK with an indefinite lifetime (Shirron, 2015). The MIRECLE temperature stability requirement at the 50 mK stage (2.5~$\mu$K rms over 10 min) is similar to that of Hitomi. The Hitomi design and temperature readout system easily meets this requirement. 

\vspace*{-0.5\baselineskip}
\subsection{Transition Edge Sensors}
\vspace*{-0.5\baselineskip}

Only a very stable detector will allow for the stability requirements needed, even with the aforementioned calibration system. TES detectors have been measured to produce white, flat and featureless noise spectra down to ~10 mHz or below. Being bolometer devices, TES can operate at any wavelength, as long as the photons can be efficiently absorbed detector pixels. Reliable coupling schemes exist for Mid-IR wavelengths, we only have to optimize them for the required wavelength range.  The Physics of TES is well understood in detail, so the performance of those devices can be reliably designed.

\vspace*{-0.8\baselineskip}
\section{Organization, Partnerships, \& Current Status}
\vspace*{-0.5\baselineskip}


The MIRECLE mission concept has leveraged the expertise of a multi-disciplinary team of exoplanet scientists and instrumentalists in order to produce the current mission concept.  The team has also benefited from cross-institutional collaboration with team members from NASA Goddard, the Space Telescope Science Institute, Johns Hopkins University, and the Applied Physics Laboratory. The novel application of TES detectors in tandem with an advanced calibration strategy required for the proposed exoplanet observations, and the assessment of the scientific capabilities of a potential mission and the required architecture was enabled through these partnerships. In preparation for a mission proposal, further partnerships with industry and university partners would be initiated.

\vspace*{-0.8\baselineskip}
\section{Schedule}
\vspace*{-0.5\baselineskip}


MIRECLE can be divided into four pieces which can be built and tested concurrently. The individual parts can then be assembled and launched with minimal full scale testing. The telescope consists of the primary mirror, secondary mirror, sunshade and supports. The testing that needs to be done is primarily optical and much of the testing can be done in room-temperature facilities.

The first instrument components to be fabricated and assembled will be the detectors, since this is typically the longest lead-time element. For MIRECLE the detector is similar to other TES detectors built and tested at
GSFC, and we can take advantage of the existing infrastructure of dewars, and readout electronics. The assembled instrument itself, consisting of the calibrator and the spectrometer, will be relatively small and so can be tested in inexpensive existing dewars with minimal modification. 

Finally the spacecraft is conventional with solar panels for power and conventional communication requirements. The pointing requirements are not severe and can take advantage of the fact that MIRECLE will be staring at a reasonably bright star, which can be used as a guide star. Thus the spacecraft bus can be sourced from an aeronautical company for a fixed price. We estimate a total of 5 years between selection and launch.

\vspace*{-0.8\baselineskip}
\section{Cost Estimates}
\vspace*{-0.5\baselineskip}



Potential cryocooler providers’ rough order of magnitude estimate of the cost of one flight cryocooler and a spare parts for this mission is on the order of \$30M.  The ADR cost estimated from parametric analysis one flight unit and spare parts is about \$20M.  A parametric telescope cost model has been developed by Phil Stahl (Stahl, et al., 2012). Using the three most important parameters: diameter of primary, D; temperature of telescope, T; and diffraction limited wavelength performance, L, and a normalizing parameter from Phil Stahl’s data base, we get the cost, C:

$C = A_sD^{1.7} T^{-0.25} L^{-0.3}  =$ \$28M  in \$2020 before integration with the rest of the observatory.

While we have not had the time/resources to study the instrument and observatory/systems cost, we estimate this mission to fit into the "Small" category (less than \$500M).

\vspace*{-0.8\baselineskip}
\section{Acknowledgements}
\vspace*{-0.5\baselineskip}

We would like to thank Tiffany Kataria and Jonathan Fortney for their contributions to the mid-IR science case.

\section{References}

\begin{thebibliography}{}
\expandafter\ifx\csname natexlab\endcsname\relax\def\natexlab#1{#1}\fi

\bibitem[{{Delrez} {et~al.}(2018){Delrez}, {Gillon}, {Queloz}, {Demory},
  {Almleaky}, {de Wit}, {Jehin}, {Triaud}, {Barkaoui}, \&
  {Burdanov}}]{Delrez2018}
{Delrez}, L., {et~al.} 2018, in Society of Photo-Optical Instrumentation
  Engineers (SPIE) Conference Series, Vol. 10700, \procspie, 107001I

\end{thebibliography}
\nobibliography{main}

\noindent\hangindent=2em Angerhausen, D., Sapers, H., Citron, R., Bergantini, A., Lutz, S., Queiroz, L. L., Alexandre, M., Araujo, A. C. V., 2013. “HABEBEE: Habitability of Eyeball-Exo-Earths.” Astrobiology, 13.

\noindent\hangindent=2em Barclay, T., Pepper, J., Quintana, E. V., 2018. “A Revised Exoplanet Yield from the Transiting Exoplanet Survey Satellite (TESS).” The Astrophysical Journal Supplement Series, 239, 2.

\noindent\hangindent=2em Boutle, I. A., Mayne, N. J., Drummond, B., Manners, J., Goyal, J., Hugo Lambert, F., Acreman, D. M., Earnshaw, P. D., 2017. “Exploring the climate of Proxima B with the Met Office Unified Model.” Astronomy \& Astrophysics, 601, A120.

\noindent\hangindent=2em Breedlove, 2014.  J. Breedlove and et al, “Testing of a two-stage 10 k turbo-brayton cryocooler for space applications,” Cryocoolers 18, 445–452 (2014).

\noindent\hangindent=2em Carone, L., Keppens, R., Decin, L., Henning, T., 2018. “Stratosphere circulation on tidally locked ExoEarths,” Monthly Notices of the Royal Astronomical Society, 473.

\noindent\hangindent=2em Crossfield, I. J. M., Hansen, B. M. S., Harrington, J., Cho, J. Y.-K., Deming, D., Menou, K., Seager, S., 2010. “A New 24 {\micron} Phase Curve for Ups Andromedae b.” The Astrophysical Journal, 723.

\noindent\hangindent=2em \bibentry{Delrez2018}

\noindent\hangindent=2em Dressing, C. D., Charbonneau, D., 2015. “The Occurrence of Potentially Habitable Planets Orbiting M Dwarfs Estimated from the Full Kepler Dataset and an Empirical Measurement of the Detection Sensitivity.” The Astrophysical Journal, 807, 45.

\noindent\hangindent=2em Driscoll, P. E., Barnes, R., 2015. “Tidal Heating of Earth-like Exoplanets around M Stars: Thermal, Magnetic, and Orbital Evolutions.” Astrobiology, 15.

\noindent\hangindent=2em Fulton, B. J., Petigura, E. A., 2018. “The California-Kepler Survey. VII. Precise Planet Radii Leveraging Gaia DR2 Reveal the Stellar Mass Dependence of the Planet Radius Gap,” The Astronomical Journal, 156, 264.

\noindent\hangindent=2em  Glaister, 2007.  Glaister, D.S., Gully, W., Ross, R.G., Jr., et al., “Ball Aerospace 4-6 K Space Cryocooler,” Cryocoolers 14, ICC Press, Boulder, CO, 2007, pp. 41-48.

\noindent\hangindent=2em Kopparapu, R. K., Wolf, E. T., Arney, G., Batalha, N. E., Haqq-Misra, J., Grimm, S. L., Heng, K., 2017. “Habitable Moist Atmospheres on Terrestrial Planets near the Inner Edge of the Habitable Zone around M Dwarfs,” The Astrophysical Journal, 845, 5.

\noindent\hangindent=2em  Olson, 2005. J.R. Olson, M. Moore, P. Champagne, E. Roth, B. Evtimov, J. Jensen, A. Collaço, T. Nast, “Development of a Space-Type 4-Stage Pulse Tube Cryocooler For Very Low Temperature”, Advances in Cryogenic Engineering 51. pp 623-631 (2005).

\noindent\hangindent=2em Pierrehumbert, R. T., 2010. “Principles of Planetary Climate.” Cambridge University Press. ISBN: 9780521865562, 2010.

\noindent\hangindent=2em Shields, A. L., Ballard, S., Johnson, J. A., 2016. “The habitability of planets orbiting M-dwarf stars.” Physics Reports, 663.

\noindent\hangindent=2em  Shirron, 2002.  Shirron P et al. 2000 Adv. Cryo. Eng. 45 pp 1629-1638.

\noindent\hangindent=2em  Shirron, 2015. Peter J. Shirron, Mark O. Kimball, Michael J. DiPirro, and Thomas G. Bialas, Performance testing of the Astro-H flight model 3-stage ADR, Physics Procedia 67 (2015), pp. 250-257.

\noindent\hangindent=2em Staguhn,J., Fixsen, D., Stevenson, K., Moseley, S.H., Sharp, E., Brown, A., Fortney, J., Hilton, G., Kataria, T., Wollack, E., 2019, "An Ultra-Stable Mid-Infrared Sensor for the Detection of Bio-Signatures by Means of Transit Spectroscopy", IEEE Aerospace Conference, DOI: 10.1109/AERO.2019.8741666

\noindent\hangindent=2em  Stahl, 2013.	H. Philip Stahl, Todd Henrichs, Alexander Luedtke and Miranda West, “Update on multi-variable parametric cost models for ground and space telescopes”, Proc. of SPIE Vol. 8442 844224-1 (2012).

\noindent\hangindent=2em Turbet, M., Leconte, J., Selsis, F., Belmont, E., Forget, F., Ribas, I., Raymond, S. N., Anglada-Escudé, G., 2016. “The habitability of Proxima Centauri b. II. Possible climates and observability.” Astronomy \& Astrophysics, 596, A112.

\noindent\hangindent=2em  Tuttle, 2017.  James Tuttle, et al., 2017 IOP Conf. Ser.: Mater. Sci. Eng. 278 012009.

\noindent\hangindent=2em Wakeford, H. R., Sing, D. K., 2015. “Transmission spectral properties of clouds for hot Jupiter exoplanets,” Astronomy \& Astrophysics, 573, A122.


\noindent\hangindent=2em Wolf, E. T., 2017. “Assessing the Habitability of the TRAPPIST-1 System Using a 3D Climate Model.” The Astrophysical Journal Letters, 839, L1.

\noindent\hangindent=2em Yang, J., Cowan, N. B., Abbot, D. S., 2013. “Stabilizing Cloud Feedback Dramatically Expands the Habitable Zone of Tidally Locked Planets,” The Astrophysical Journal Letters, 771, L45.

\noindent\hangindent=2em Zahnle, K. J., Marley, M. S., 2014. “Methane, Carbon Monoxide, and Ammonia in Brown Dwarfs and Self-Luminous Giant Planets,” The Astrophysical Journal, 797, 41.

\end{document}